\documentclass[debug,overfull,figures,info,cite]{epl}
\usepackage{graphics}
\title{Prewetting transition on a weakly disordered substrate : \\
evidence for a creeping film dynamics}%
\shorttitle{Prewetting transition on a weakly disordered\ldots}
\author{X. M\"uller, J. Dupont-Roc }
\institute{Laboratoire Kastler Brossel \thanks{Unit\'e
 mixte de recherche du CNRS, de l'Ecole Normale Sup\'erieure et de l'Universit\'e Pierre et
Marie Curie},  Ecole Normale Sup\'{e}rieure, \\ %
 24 rue Lhomond, F75231 Paris cedex 05, France}%

\begin{document}

\maketitle

\begin{abstract}
We present the first microscopic images of the prewetting transition of a liquid film on a solid
surface. Pictures of the local coverage map of a helium film on a cesium metal surface are taken
while the temperature is raised through the transition.  The film edge is found to advance at
constant temperature by successive avalanches in a creep motion with a macroscopic correlation
length. The creep velocity varies strongly in a narrow temperature range. The retreat motion is
obtained only at much lower temperature, conforming to the strong hysteresis observed for
prewetting transition on a disordered surface. Prewetting transition on such disordered surfaces
appears to give rise to dynamical phenomena similar to what is observed for domain wall motions in
2D magnets.
\end{abstract}
\section{Introduction}
Over the past twenty years, prewetting transition has attracted a lasting interest since its first
consideration by Ebner, Saam and Cahn\cite{ebner77,cahn77,ebner87}.  Considering an equilibrium
between two fluid phases, the prewetting transition on a surface is simply the continuation off
coexistence of the wetting transition. The surface coverage corresponding to the wetted state is
simply finite rather than infinite. Apart from very special matching
conditions\cite{dietrich,indekeu} for which critical wetting transition has indeed been
observed\cite{meunier2}, first order transitions are expected, resulting from long range Van der
Waals forces between the surface and the various fluid components\cite{cahn77,ebner87,dietrich}. In
the simple case of a liquid/vapor phase equilibrium in presence of a surface, the surface coverage
changes from a thin coverage to a thick liquid film. Experimental observations of these transitions
on various systems have been made over the last
decade\cite{hallock1,taborek1,meunier1,demolder1,demolder2,mistura,taborek7,reinelt1,reinelt2,reinelt3,reinelt4,yao1,yao2,kozhevnikov1,wynblatt,lucht2}.
A particular goal of experimentalists has been to check the first order nature of the transition,
to determine the prewetting line in the $T$~-~$\Delta\mu$ plane and to locate the prewetting
critical point. As a matter of fact, experiments have raised further questions about hysteresis
phenomena, about the consequences of surface disorder, about the nucleation of the new phase at the
transition and its subsequent growth. Presently there is no unique interpretation of the various
experimental observations of prewetting transitions, especially at low temperatures. To our best
knowledge, nobody has ever provided a clear experimental evidence for the coexistence at the
prewetting transition of areas with thick and thin coverages, separated by an interface which can
be moved in a \emph{reversible} way, as in a \emph{bona fide} first order transition such as the
liquid/vapor one. The transition reversibility has not always been tested. When it has been looked
for, hysteresis has been found quite
systematically\cite{hallock1,taborek1,meunier1,demolder1,demolder2,taborek7,reinelt1,reinelt2,reinelt3,reinelt4}.
This hysteresis has received diverging interpretations. For some, it reveals a nucleation barrier
which can be overcome only by over-passing the equilibrium line and approaching the spinodal
line\cite{meunier1,bonn1,herminghaus}. Such a barrier is indeed expected for instance for the
dewetting branch of the transition when the initial surface is completely covered by a thick film
and that the thin film phase is to be nucleated. For others, hysteresis is related to surface
disorder, such as local variations of the surface binding properties. They are equivalent to a
random field coupled to the surface coverage, which is the order parameter. This random field is
expected to destroy the first order transition. Dynamical transitions may occur for the invasion or
drying of the surface\cite{robbins1,blossey1,blossey2}. Then experimental data are to be related
to numerous studies about growth mechanisms of 2D phases\cite{leger92,cazabat98,lemerle98}. Both
interpretations may be true, depending of the substrate. While in demixing liquids, the
liquid-vapor interface play the role of a perfectly smooth substrate for which the first
interpretation is relevant, solid surfaces are thought to exhibit frozen disorder. Thus detailed
experimental data are still desired to justify one or the other interpretation. In many experiments
only the overall coverage of the surface, or another surface averaged, quantity is measured. Hence
a detailed view of the transition is not accessible. Conversely some
experiments\cite{reinelt1,reinelt2,reinelt3} have provided detailed images or movies of spreading
films, but clearly out of equilibrium. In this letter, we report the first observation of the
coverage map during a prewetting transition. The system used is liquid helium on a cesium metal
surface, which includes some frozen disorder. The invasion of the surface by the thick film at the
transition is found to take place through a creeping motion of the thick film edge.

\section{Experimental procedure}
Cesium metal surfaces are prepared by cold deposition from SAES Getters dispensers on a gold
evaporated mirror in conditions similar to other
experiments\cite{hallock1,taborek5,reinelt1,mistura,rolley1}. Some care is taken to avoid
impurities on the substrate~: dispensers, as well as the cell and the pumping tubes, are outgazed
at room temperature. To improve its uniformity, the cesium film is
 deposited from two dispensers at $\pm 45^\circ$ with respect to the surface normal. It extends
over a $2.5\times 4\ \mathrm{mm}^2$ rectangle determined by a mask located 0.4~mm above the center
of the gold mirror. For the experiments reported here, the cesium layer was 40$\pm$5~nm thick. Last
layers were deposited at low rate (0.3 nm/min). Differently prepared cesium substrates have also
been used in wetting experiments with liquid helium\cite{wyatt1,wyatt2}. For these surfaces,
contact angles and hysteresis properties for the wetting transition have been reported to be
markedly different from those obtained with the type of surface used here. Prewetting transitions
on such substrates have not however been investigated.

Cesium thickness and helium coverage are measured through an improved Nomarski microscope adapted
to low temperatures\cite{gleyzes97,muller1,muller2}. It provides a map of the local gradient along
a given direction of the local surface height. The corresponding image looks like a side
illuminated picture of the true surface landscape with a nanometer height scale, and with the
microscope resolution (about 4~$\mu$m in our case). If a transparent coverage is added on the
surface, it produces an optical retardation which appears as an additional negative height. Film
edges appear as bright or dark lines. With a 6~s integration time, the microscope is sensitive to
optical thickness differences as small as 0.05~nm over a distance of 4~$\mu$m. That is sufficient
to detect 10~nm thick helium films which correspond to a 0.26~nm optical retardation ($n_{\rm liq.
He}-1 = 0.026$).

 Below 1.8~K, liquid helium does not wet the cesium surface. When filling the cell with helium liquid
 at low temperatures (typically 1.4~K), the vertical cesium mirror surface does not get
wetted. When the liquid level is raised onto the surface, it makes a non-zero advancing contact
angle. When afterwards the level is lowered by pumping part of the helium out of the cell, the
contact angle becomes zero and a metastable thick film remains pinned on the surface. The film
thickness is determined by its altitude above the liquid level which can be varied from 1~mm to
20~mm. The corresponding measured thicknesses are 70~nm and 20~nm. These values are consistent with
those computed from the cesium/helium non-retarded Van der Waals force\cite{vidali1}. The mirror
surface around the small rectangle covered with cesium is of course also covered by a wetting
helium film. When temperature is raised, the prewetting transition temperature is eventually met
and the helium film should cover the originally dry area. Afterwards, lowering the temperature
should cause dewetting. Because the initial state contains both thin and thick phases in contact,
nucleation metastability is avoided. Interface pinning by the disorder is the only hysteresis
mechanism at work in this situation.

\section{Prewetting transition through thick film invasion}
\begin{figure}
\threeimages[scale=0.5]{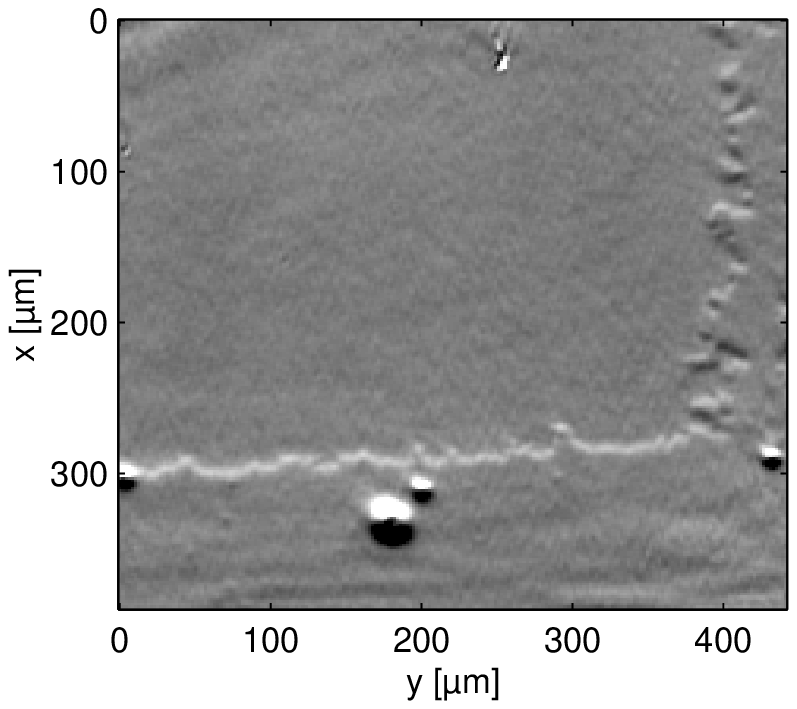}{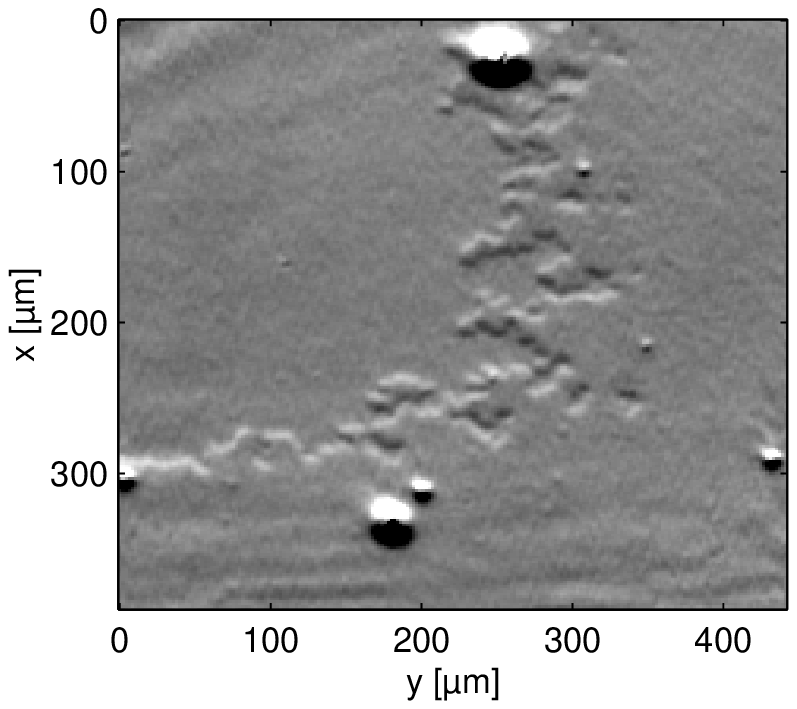}{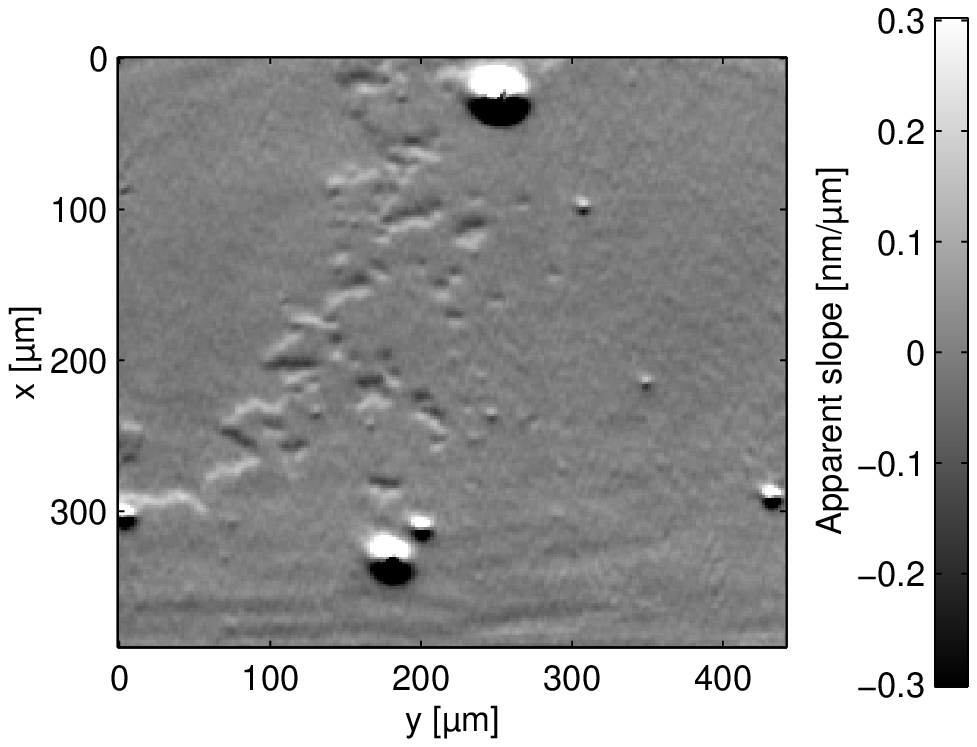}%
\caption{Invasion of the cesium surface by the helium film at 1.817~K. Pictures are taken
respectively 4, 70 and 140~min after reaching this temperature. The gray scale is related to the
film optical thickness gradient along the $x$-direction. The actual film thickness is about $50\pm
8$~nm. Microscopic dust grains give rise to some local thickening of the film appearing as `helium
hills' as soon as they are reached by the film front.}%
\label{avancee}
\end{figure}

In a first run, temperature is increased slowly, typically at a rate less than 1~mK/min, until a
change is observed in the coverage. Then temperature is stabilized during the time of the
experiment within a few mK. In subsequent runs, temperature is set at different values around this
initial one. It is observed that the straight line of the initial film border becomes irregular
because it starts to move at a few places. Then the film advances over the dry area with a slow
creeping motion by successive avalanches. Intermediate states of the surface invasion at 1.817 K
are shown in Figure \ref{avancee}. Because the observed area is near the right edge of the cesium
rectangle, part of invasion comes also from the right. This explains the overall L-shaped film
front. The contrast of the film edge does not change significantly, indicating a constant
thickness.

\begin{figure}
\twofigures[scale=0.33]{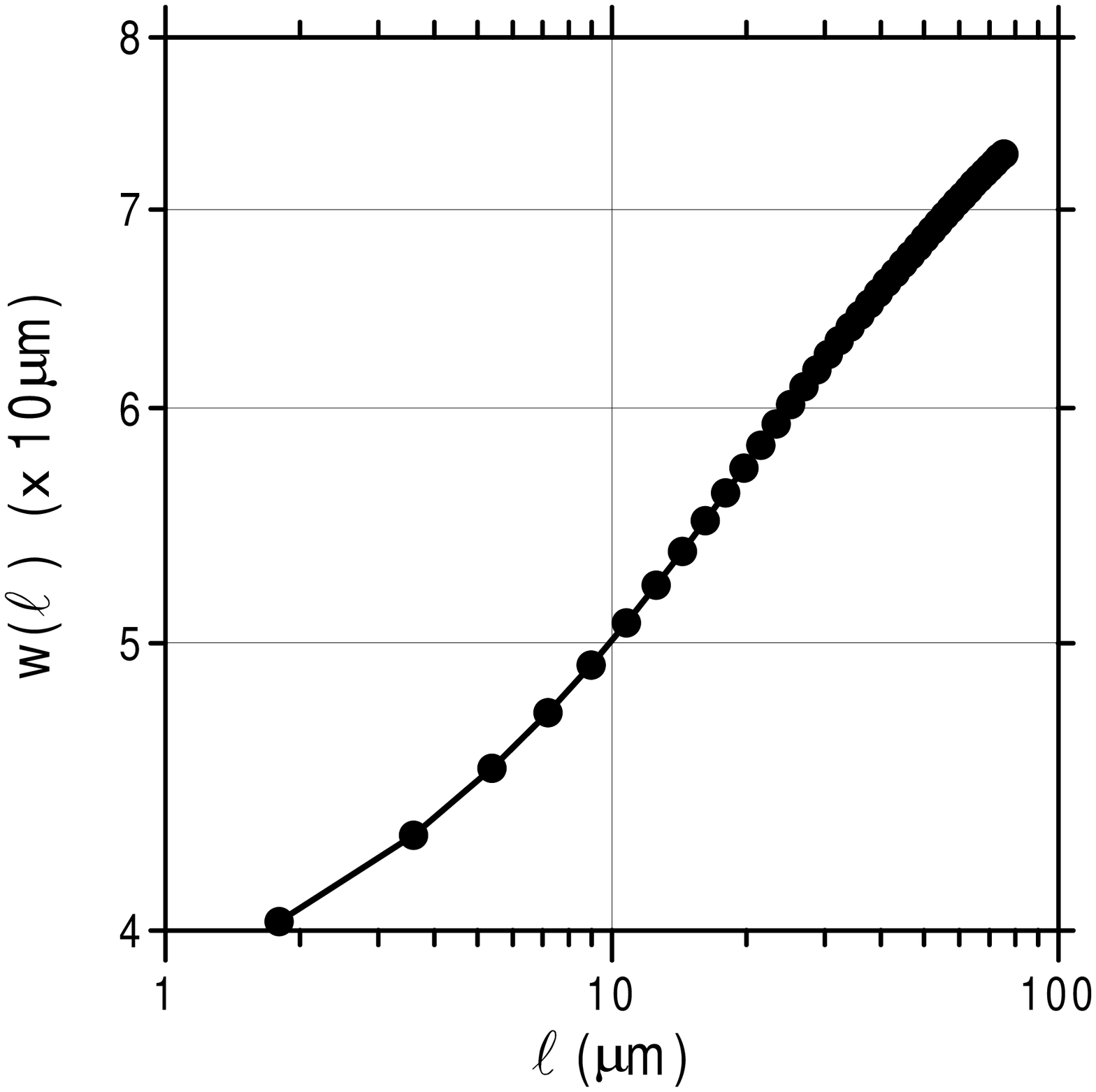}{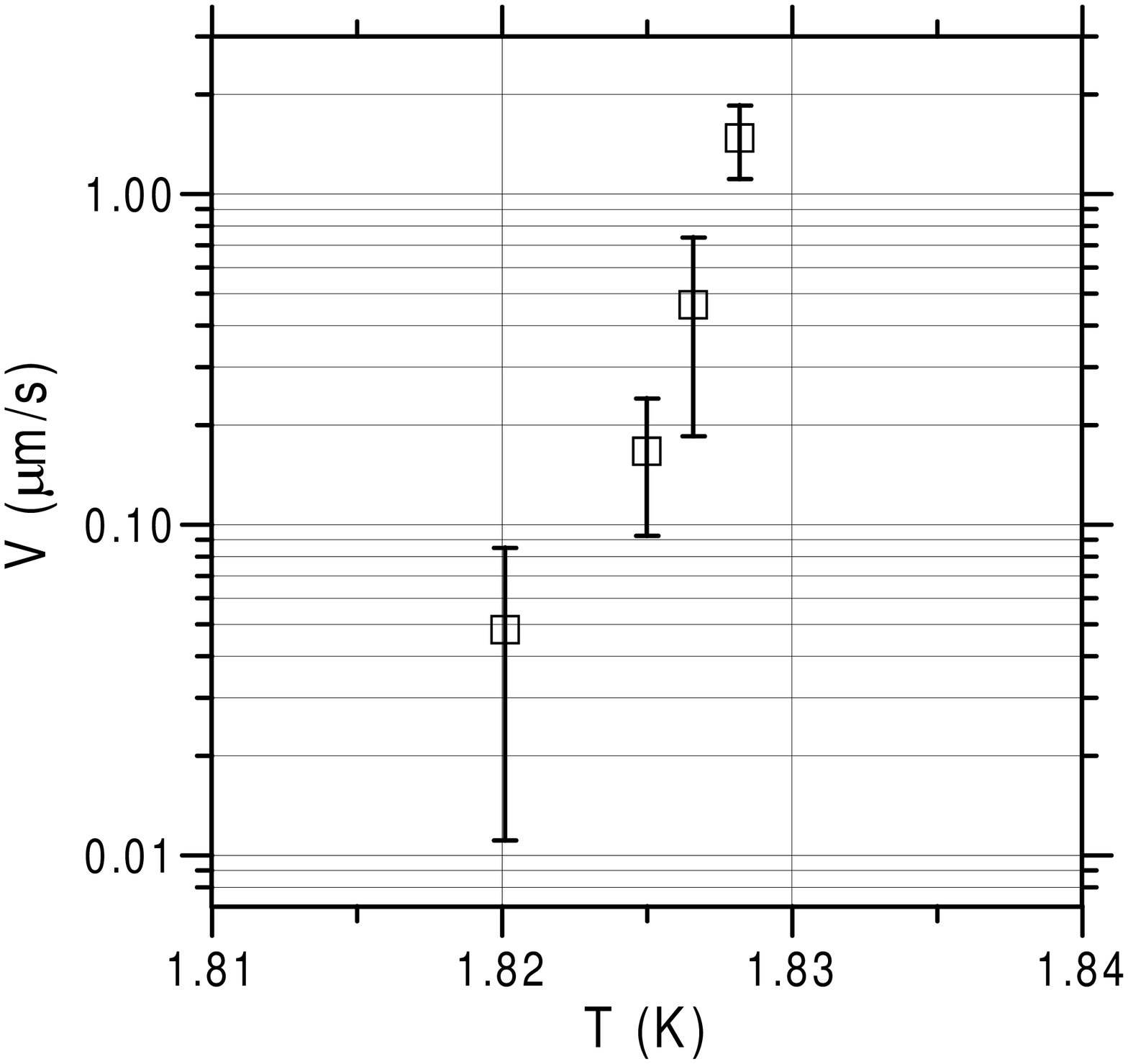}
\caption{$w(l)$, r.m.s front position fluctuations on length $l$ along a line approximating the front. %
The average is taken over several successive positions of the film border and on several %
experimental runs. Non vanishing values at short distances are due to overhangs. }%
\label{correlation}%
\caption{Creep velocity as a function of temperature for $50\pm 8$~nm thick films}%
\label{vitesse}
\end{figure}

The morphology of the film border is characterized by a macroscopic correlation length, clearly
visible in Figure \ref{avancee}, much larger than typical lengths characterizing the cesium layer
such as its thickness 40~nm, or its morphological length scale which is probably smaller. This is a
clear indication of a cooperative (i.e. weak) pinning situation for the film edge. To quantify the
correlation length, one considers the r.m.s deviation of the film edge position $w(l)$ taken over a
length $l$ along the mean front direction. It is plotted versus $l$ in Figure \ref{correlation}. At
short distance, the plot starts from a non-zero value due to existing overhangs and dry holes
inside the film.  $w(l)$ increases over long distances. Cross-over between the two regimes occurs
for the correlation length $l_c \simeq 10$~$\mu$m and  a front depth $w(l_c) = 50$~$\mu$m. For zero
temperature models, Robbins \emph{et al.} point out that this cross-over prevents from getting the
roughness exponent 0.5 expected for a self-affine growth\cite{robbins1}. We are not aware of
predictions for $w(l)$ taking into account both finite temperatures and important contributions of
overhangs.

 No nucleation of wet patches in front of the film seems to occur. It is unlikely that wet parts
smaller than the microscope resolution exist. In particular, some defects (presumably submicronic
dust grains) give rise to `helium hills' inside the film by a local thickening, as can be seen in
Figure \ref{avancee}. These `hills' appears only when the film edge reaches them. This suggests
that the surface ahead the film is indeed free of thick liquid film.

After the cesium surface is completely invaded, the cell is cooled down as much as possible.
Partial dewetting was observed around 1.23~K. The film split into two parts separated by a dry
horizontal line, then one of this part completely disappeared.  Dewetting, as wetting, was
anisotropic. As anticipated from previous studies, the film edge dynamics is strongly hysteretic.

The film advances at constant temperature by successive avalanches with sizes on the order of
$l_c$. The advance velocity is defined as the wet area increase between two images divided by the
time interval and the average front length. For comparison, velocities are always measured on the
same area of the surface. Two facts hints at a thermally activated process. First, the dry holes
inside the film fill up after a while. This means that no part of the border is definitely pinned.
They have a chance to move if one waits for a sufficiently long time. A probabilistic mechanism is
at work to get over the higher barriers, probably thermal fluctuations since intentionally produced
vibrations have no effect. Secondly the film creep velocity increases strongly with temperature as
shown in Figure \ref{vitesse}. Higher temperatures indeed increase the spreading forces compared to
the pinning ones. This lowers the barriers and makes avalanches more probable. Film dynamics
exhibits also some peculiarities. The advance is much faster from the right to the left than
upwards, and the speed of the front motion somewhat decreases while it sweeps across the picture
area. This may be due to a slight non uniformity and anisotropy of the cesium layer related to the
deposition geometry, but this remains to be checked.

To summarize, in the case presented here, prewetting transition takes place through the
displacement of the thick film border by a random, probably thermally activated process, with a
macroscopic correlation length indicating a cooperative pinning. The creep velocity changes from an
vanishing low value to observable ones on a very small temperature interval, around what appears as
a unbinding threshold. The inverse process, the drying transition, occurs through a similar process
significantly shifted to lower temperatures.

\section{Discussion}
These properties strongly suggests that the normal first order prewetting transition is hindered by
the pinning of the thick film border by the substrate disorder. The thick film advance or retreat
takes place only when the spreading force is sufficiently strong to overcome the corresponding
barriers. The observed transition is thus related to the film border dynamics. This is consistent
with the observation of different advancing and receding angles on similar surfaces below the
wetting temperature\cite{rolley97,ross97,phillips98}. The fact that no nucleation ahead of the
front is observed means that the pinning is not strong enough to bring the observed dynamical
transition up to the spinodal line of the first order wetting transition\cite{bonn94}. The growth
itself is nearly compact. These features are in contrast with the observations in a previous
experiment presented in \cite{muller3}. With thicker cesium layers obtained at higher deposition
rates, no motion of the film edge was observable. Only a continuous increase of the coverage ahead
of the front was detected when temperature was increased above a given threshold. This was
interpreted as the growth of a  film non compact at a scale below the microscope resolution. The
substrate disorder was probably more important, and thus the film correlation length much shorter.
It is also interesting to remark that the wetting threshold was somewhat higher in temperature
(1.9~K instead of 1.82~K here) and that dewetting was not observable, suggesting an overall larger
hysteresis in term of spreading force. A similar case was recently studied for cesium surfaces with
a controlled large scale disorder\cite{prevost00}. The correlation length of the film border is
then closer to the disorder length scale indicating a strong pinning case.

Concerning the dynamics, the motion of a line in a disordered 2D medium has been described by a
thermal activation law for the velocity, with a stretched exponential form\cite{rolley1,chauve00}~:
$v\sim \exp[-Cf^{-\mu}/T ]$ where $C$ is a constant, $f$ is the driving force per unit length of
line and $\mu$ is an exponent depending the system dimensionality and disorder. Such a law has been
verified recently for the triple line motion \cite{rolley1}, and for domain wall motion in 2D
magnets\cite{lemerle98} with $\mu =1/4$. In the present case, the system is in the same
universality class as the 2D random field Ising model so that $\mu$ is expected to take the value
1. The range of velocity observed is however not wide enough to check this prediction.

The spreading of a helium prewetting film was already observed by Reinelt \emph{et
al.}\cite{reinelt1}. A droplet falling on an inclined surface was accompanied by a spreading helium
film, 30~nm thick. Its spreading velocity was about 0.5~cm/s at the temperature 1.85K.  This large
velocity may indicate a larger temperature difference from the threshold. It is remarkable, but
probably a coincidence that linear extrapolation of the data of Figure \ref{vitesse} to 1.85~K
yields the same order of magnitude. Note that with such a thermally activated law for the velocity
the temperature at which the prewetting transition is actually observed depends somewhat on the
observation time. Changing this time from 1~s to 4~hour shifts the observed prewetting temperature
only from 1.85~K to 1.82~K. Hence it is tempting to suggest that the situations appearing as
metastable in the temperature range between the observed prewetting temperature and the dewetting
one actually require exponentially large and unrealistic observation times to reach equilibrium.
Then the threshold at which the invasion begins does not correspond to a real dynamical transition
as in zero temperature models\cite{blossey1,blossey2}. It is rather a cross-over between a low
temperature regime with large and rare avalanches to a high temperature regime, not reached in this
experiment, with an overall motion of the border line limited by some dissipation mechanism.

\section{Conclusion}
The observations presented show clearly how the frozen disorder of the surface interfere with the
prewetting transition in the weak disorder situation. Collective pinning of the border line between
the thick and the thin film phases prevents the transition from taking place through its
displacement in favor of one or the other of the two phases. Experimental observations of the
transition correspond then to the unbinding of the line at a finite spreading power. This relates
prewetting transitions on disordered substrates to other phenomena exhibiting domain wall motions
in two dimensions, such as 2D magnetism.

\acknowledgments We are indebted to E.~Rolley, A.~Prevost  and C.~Guthmann for many helpful
discussions.

\end{document}